# Interstitial dual-mode ultrasound with a 3-mm MR-compatible catheter for image-guided HIFU and directional *in-vitro* tissue ablations

Thomas Biscaldi[1], Romain L'Huillier[2,1], Laurent Milot[2,1], W. Apoutou N'Djin[1], *Member, IEEE*

[1] LabTAU, INSERM, Centre Léon Bérard, Université Claude Bernard Lyon 1, F-69003, LYON, France
[2] Department of Diagnostic and Interventional Radiology, Hôpital Edouard Herriot, Hospices Civils de Lyon, University of Lyon, 69003 Lyon, France

e-mails: thomas.biscaldi@inserm.fr, romain.lhuillier@chu-lyon.fr, laurent.milot@chu-lyon.fr, apoutou.ndjin@inserm.fr

***Abstract*** - Current interstitial techniques of tumor ablation face challenges that ultrasound technologies could meet. The ablation radius and directionality of the ultrasound beam could improve the efficiency and precision. Here, a 9-gauge MR-compatible dual-mode ultrasound catheter prototype was experimentally evaluated for Ultrasound Image-guided High Intensity Focused Ultrasound (USgHIFU) conformal ablations. The prototype consisted of 64 piezocomposite linear array elements and was driven by an open research programmable dual-mode ultrasound platform. After verifying the US-image guidance capabilities of the prototype, the HIFU output performances (dynamic focusing and HIFU intensities) were quantitatively characterized, together with the associated 3D HIFU-induced thermal heating in tissue phantoms (using MR thermometry). Finally, the ability to produce robustly HIFU-induced thermal ablations in in-vitro liver was studied experimentally and compared to numerical modeling. Investigations of several HIFU dynamic focusing allowed overcoming the challenges of miniaturizing the device: mono-focal focusing maximized deep energy deposition, while multi-focal strategies eliminated grating lobes. The linear-array design of the prototype made it possible to produce interstitial ultrasound images of tissue and tumor mimics in situ. Multi-focal pressure fields were generated without grating lobes and transducer surface intensities reached up to $I_{sapa}$ =14 W·cm$^{-2}$. Seventeen elementary thermal ablations were performed in vitro. Rotation of the catheter proved the directionality of ablation, sparing non-targeted tissue. This experimental proof of concept demonstrates the feasibility of treating volumes comparable to those of primary solid tumors with a miniaturized USgHIFU catheter whose dimensions are close to those of tools traditionally used in interventional radiology, while offering new functionalities.

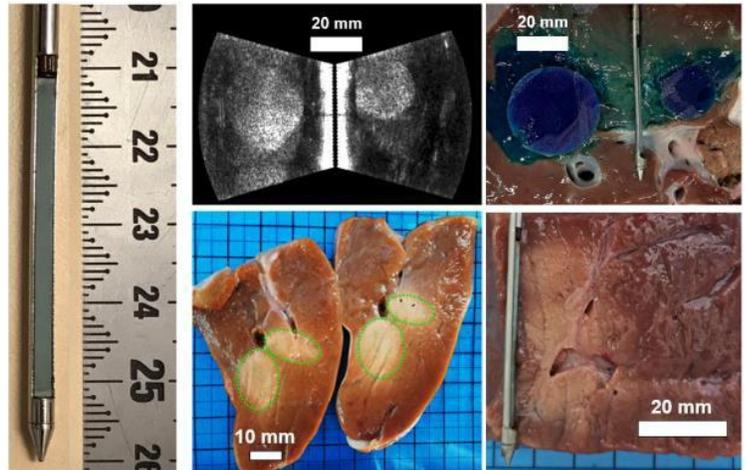

*Highlights*

- *This study presents a 3 mm diameter dual-mode interstitial ultrasound catheter capable of operating in both High Intensity Focused Ultrasound (HIFU) therapy mode and ultrasound (US) imaging mode with the same transducer.*
- *Directional radial thermal ablations up to 26 mm deep were achieved with multiple HIFU focusing strategies, while target tumor-mimics were accurately detected with B-mode US imaging.*
- *This type of medical device would overcome some of the remaining challenges encountered in interventional radiology, offering extensive directional ablations and in situ imaging using a single needle.*

## I. Introduction

Minimally invasive tissue ablation techniques for localized cancer treatment offer the advantage of making only small incisions in the body, reducing the risk of complications and recovery time and can be offered to patients non eligible to surgery. Interstitial thermal ablation procedures rely on the use of interstitial catheters that locally modify tissue temperature in such a way as to induce irreversible ablations. In today's clinical context, the most commonly used techniques inducing tissue heating are radio-frequency ablation (RFA) [1,2], microwave ablation (MWA) [3–5] and laser interstitial thermal therapy (LITT) [6–8]. Cryo-ablation (CRA) is based on tissue cooling [9,10]. Irreversible electroporation (IE) is also becoming an increasingly important treatment option [11,12].

Current interstitial treatment techniques work mainly by thermal diffusion which is an omnidirectional phenomenon [13]. Consequently, the shape of the final ablation zones produced by current standard devices on the market cannot be adapted in 3D to a given target region, which makes these treatment strategies not conformal. It remains possible to approach conformal treatment by inserting several needles into the patient [14]. However, the technique becomes much more complex for the practitioner, increasing significantly the procedure duration and the risk of bleeding [15]. Moreover, it remains difficult to perform conformal ablation in 3D without real-time guidance and monitoring strategies. In addition, ablations induced by thermal diffusion are highly dependent on perfusion, tissue type and the overall environment in the local target zone. All these parameters increase the difficulty of spatially controlling the treated area and lead to under or over- treatment and narrow safety margins, increasing the risk of residual tumor and recurrence [16].

For instance, in the case of hepatocellular carcinoma (HCC) the aim is to treat the tumor with sufficient margins, while sparing as much non-tumor liver as possible. When the tumor is located close to a sensitive or at-risk part of the body (e.g. digestive tractor bile ducts), it is vital not to damage these tissues, so as not to cause severe complications for the patient [17]. Oftentimes, to overcome these issues, very complex strategies, involving aggressive hydrodissection or balloon insertion are utilized in expert centers. It is essential to ensure that the ablation performed does not exceed the at risk margin set by the practitioner. A directional ablation technique would allow orienting the treatment in a given tissue region. Ultimately, such technique would allow prescribing conformal treatments with multiple controlled directional ablations adapting to the 3D targeted tumor shapes, while preserving as much healthy tissue as possible and preserving at risk structures.

To date, conformal ablative ultrasound therapies have been successfully proposed with the assistance of MR thermometry monitoring in the context of transurethral treatment of localized prostate cancers [18,19]. But these techniques cannot be implemented in many medical applications where dedicated MRI is not available. For instance, the management of HCC in interventional radiology is almost exclusively performed in MR free operative room environments [20]. Extracorporeal real- time 2D ultrasound imaging is standardly used by the practitioner to guide the whole procedure. In the most advanced interventional radiology rooms, the procedure can be assisted by the use of Computed Tomography (CT) imaging and medical robotics [21,22]. Image fusion of extracorporeal real- time 2D US imaging with extracorporeal 3D CT imaging within a virtual navigation environment thus provides pseudo-real-time 3D guidance of the interventional procedure [23]. But it is currently impossible to obtain images directly from the inside of the tumor that would provide additional information in the referential of the therapeutic interstitial needle. Treatments are thus currently monitored qualitatively (echogenicity) in the 2D plan of the extracorporeal US image positioned by the practitioner depending on the target tumor position, needle insertion orientation, itself guided by the anatomical structure configurations of each patient. There is therefore a clear interest to develop a treatment strategy and associated technology that would enable both conformal therapies and conformal therapy guidance by imaging, while remaining fully compatible use with current clinical practices and environments.

In the context of interstitial ultrasound therapies, the main objectives and challenges are to ensure extended, accurate and homogeneous deposition of the ultrasound energy into target tissues in front of the device's active surface. This require developing both: i) therapeutic US strategies for spatial- temporal control the US energy emission ii) real-time imaging strategies for guiding/monitoring the biological effects induced by the US energy deposition within target tissues. In the past three decades, many interstitial ultrasound devices have been developed, associated to multiple configurations in order to emit high US energy interstitially. They can be classified in two main categories depending on whether ultrasound waves are spatially focused or not: High Intensity Collimated Ultrasound (HICU) and High Intensity Focused Ultrasound (HIFU). In the late 90s, encouraging proof-of-concept studies were carried out [24–26], but technical limitations (no focusing or imaging) prevented the development of such prototypes in the clinic. In the early 2000s, low-density, single-mode multi-element HICU prototypes comprising between 4 and 10 elements were developed [18,19,27]. But these strategies, not compatible with dual-mode ultrasound imaging, require MRI for treatment guidance. In addition, the dimensions of these catheters (e.g. Ø ~ 6 mm or ~ 6 G) were exclusively compatible with a practical clinical use in the field of Natural Orifice Transluminal Endoscopic Surgery (NOTES), more specifically for transurethral approaches [18,19]. For interstitial approaches, acceptable catheter dimensions must be considered in regard to existing interstitial strategies (RFA, MWA, LITT, CRA: Ø <= 9 G or ~ 3.3 mm). The efficacy of HICU strategies thus becomes challenged by the rapid divergence of the emitted US energy, as catheters miniaturization requires decreasing drastically transducers elevations [28].

Interstitial HIFU strategies can then overcome these limitations by allowing concentration and deposition of high US energy over larger radial distance with miniaturized catheters. HIFU can be achieved either with geometrically and/or dynamically focused ultrasound. In the late 2000s, low- density (5 to 10 elements) focused multi-element HIFU catheter prototypes were developed, with decreased dimensions (Ø = 4 mm or ~ 8 G) [29–31]. In parallel, the potential of ultrasound dual-modality has also been successfully proposed with these types of catheters to provide both HIFU therapy and US guidance (USgHIFU). Especially, the feasibility to reconstruct 2D B-mode US imaging by mechanically scanning the tissues with prototype rotations was demonstrated.

Finally, the development of HIFU transducers arrays allows to dynamically focused the ultrasound wave with the advantage to control and modify the position of the focus electronically. When dual-mode, these transducers allow achieving US image guidance of HIFU therapies in real-time. The most standard transducer design is a planar linear-array capable of forming 2D B-mode US imaging in real-time and of dynamically focusing high US energy for thermal therapy. However, the number and size of the elements in the array are limited by many technological challenges in a small catheter: i) mechanical integration (maximum number of elements which can be fabricated within the transducer, fixation and constrain); ii) electrical integration (maximum number and miniaturization of independent connections in a small catheter); iii) subsequent thermal and acoustical behaviors (increased self-heating and decreased ultrasound output due to mechanical and electrical losses magnified by miniaturization, efficacy of a miniature cooling system, integration of miniature sensors to monitor the transducer catheter). In the 2000s, a high-density (32 elements) dual-mode multi-element rigid catheter prototype with improved dimensions (Ø = 3 mm or ~ 9 G) showed promising results for liver tissues HIFU ablations, but were never followed up [32,33]. In the 2010s, a 64-element linear-array intracardiac flexible catheter prototype with similar diameter was developed to generate USgHIFU ablations in the atrium with precise juxtapositions of millimetric elementary ablations [34]. More recently, a team has been working on an endoluminal cylindrical applicator [35]. To date, however, no interstitial USgHIFU catheter compatible with above mentioned requirements of interventional radiology has been proved robust enough to induce conformal thermal ablations of tissue volumes compatible with HCC treatments. Consequently, to this date, no interstitial HIFU systems are available on the market.

In the field of interventional radiology, however, the development of MR-independent (while –compatible), miniature (≤ 9 gauges), dual-mode ultrasound catheters capable of performing both *in-situ* imaging and therapy remains mandatory to overcome the challenges identified above in proposing interstitial percutaneous conformal therapies.

Initial work conducted by our team led to the development of an ultrasound navigation-guided robot-assisted HIFU platform which was first evaluated in the context of interstitial conformal ultrasound therapies [36]. First, this platform enabled 3D ultrasound reconstructions of various tumor-mimic volumes embedded in *in-vitro* liver tissue samples using a first version of USgHIFU catheter (V1) controlled by a robotic arm (real-time 2D US imaging + catheter mechanical rotations/translations). Second, these tumor volume reconstructions were used to validate 3D planning of conformal HIFU treatments in numerical modeling. However, no proof of concept of HIFU treatment could be demonstrated experimentally. In addition, the relatively small length of the linear-array transducer meant that translations were required to image and treat tumor volumes with heights greater than 13 mm, by dividing the tumor volume into several stages. If the concept of interstitial 3D conformal therapies using a combination of HIFU dynamic focusing and HIFU mechanical scanning was demonstrated with a 3-mm catheter, the resulting processing times was penalized by the treatment repetition on multiple stages.

In the presented study, a second version of dual-mode USgHIFU catheter prototype (V2) has been developed with several improvements and compromises considering all these challenges. A first major evolution consisted in designing a 64- element linear-array transducer with ultrasound elements pitches larger than the ultrasound wavelength for building a 3- mm miniature catheter long enough to allow imaging and treating a tissue target (e.g. solid HCC tumor) with a single catheter insertion. A second major evolution consisted in making the prototype MR-compatible in order to allow 3D MR- thermometry characterization of the HIFU performances, while also developing a tool compatible with all above mentioned imaging guidance techniques. After reporting the technical feasibility of this development, the preliminary aim of the study was to verify the ability of the prototype to generate *in-situ* US images to guide HIFU treatments, despite the trade-off made in the design of its transducer. The main aim of the study was then to provide the first experimental proof that such a catheter can robustly generate directional HIFU ablations over tissue radii similar to those of large-diameter solid tumors (e.g. 4-6 cm- diameter HCC) with a single-needle interventional radiology procedure. Especially, an original concept of interstitial dynamic HIFU multi-focusing was proposed to overcome above mentioned technical (element pitch) and physical (ultrasound wavelength) limitations. The focusing and ablative performances of several configurations (mono-, bi-, and quadri- focal) were evaluated experimentally and compared to plane wave emissions. Finally, the mid-term robustness of this prototype was evaluated over multiple experimental HIFU trials to determine if this technology can overcome persistent limitations associated to miniaturization of HIFU catheters.

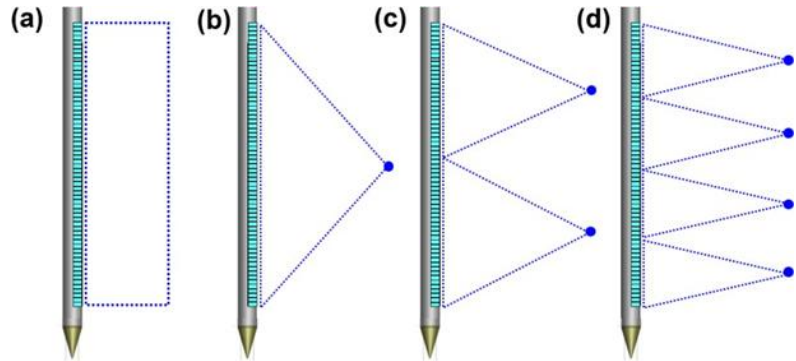

**Fig. 1**. Interstitial dynamic multi-focusing strategies.: Configurations tested in the study. a) Unfocused plane wave. b) Mono-focal wave. c) Bi-focal wave. d) Quadri-focal wave.

## II. Materials And Methods

### a. Ultrasound multi-focusing for interstitial USgHIFU ultrasound therapies: principle

In this study, several strategies have been proposed and evaluated for achieving USgHIFU with a dual-mode miniaturize interstitial catheter (Ø ≤ 3 mm or ~ 9 G), whose ultrasound elements were larger than the wavelength. The strategies to acquire B-mode US images with linear arrays are extensively described in the literature (line by line sequential acquisitions, plane or diverging waves simultaneous acquisitions, compound acquisitions).

No studies however report optimization of interstitial HIFU heat deposition from linear-arrays in these conditions. For any linear-array transducer whose pitch is significantly greater than its wavelength (phased-linear arrays), it is possible to solve the problems inherent in grating lobes by using several small groups of elements. Teams have already worked on the concept of multifocusing [37–39] and, each group of elements can focus HIFU energy on a part of the target tissues (e.g. tissues facing them). In this case, several focal points can be distributed over the target medium facing the transducer, in order to dynamically orientate and concentrate the HIFU energy (**Fig. 1**).

### b. Dual-mode US imaging & HIFU multi-focusing for conformal therapies

Then, the concept of interstitial dual-mode US-Navigation- guided HIFU assisted by medical robotics is proposed to achieve 3D conformal ablation therapies of solid tumors (e.g HCC) [36]. The procedure follows a workflow including: i) the dual-mode USgHIFU catheter insertion, ii) 2D US imaging of the target tissue, iii) 3D US image reconstruction with virtual navigation and robotics, iv) target tissue 3D segmentation, v) semi-automatic 3D HIFU treatment planning, vi) Conformal HIFU treatment in 3D, vii) monitoring during the thermal ablation induction, and viii) ablation assessment after the treatment. In the present study, a significant improvement over the previous study is proposed in the design of a new prototype. The linear array is long enough to allow imaging and treating the entire target tumor from one single catheter insertion (no additional translation) and a single rotation (**Fig. 2**).

The catheter can be implanted at the center of a target region (e.g. single large solid tumor, 4-6 cm in diameter) or outside and in the vicinity of one/several target regions (e.g. smaller solid tumors, < 3 cm in diameters). The catheter can then be rotated to scan and/or treat all tissues within the radius of action of the dual-mode linear-array transducer. For the rest of the study, 2 different spatial referential were defined. The first one is a fixed cylindrical r, h, θ referential, with r being in the radial direction of propagation of the ultrasound, h being aligned to the catheter length and θ being the catheter rotation angle over the h axis. The second one is a mobile Cartesian x, y, z referential. This referential linked to the catheter is rotating in the cylindrical r, h, θ referential, z and x axes corresponded respectively to the r ad h axes, whereas the z axis rotated to remain perpendicular to the r axis of propagation of ultrasound (**Fig 2**).

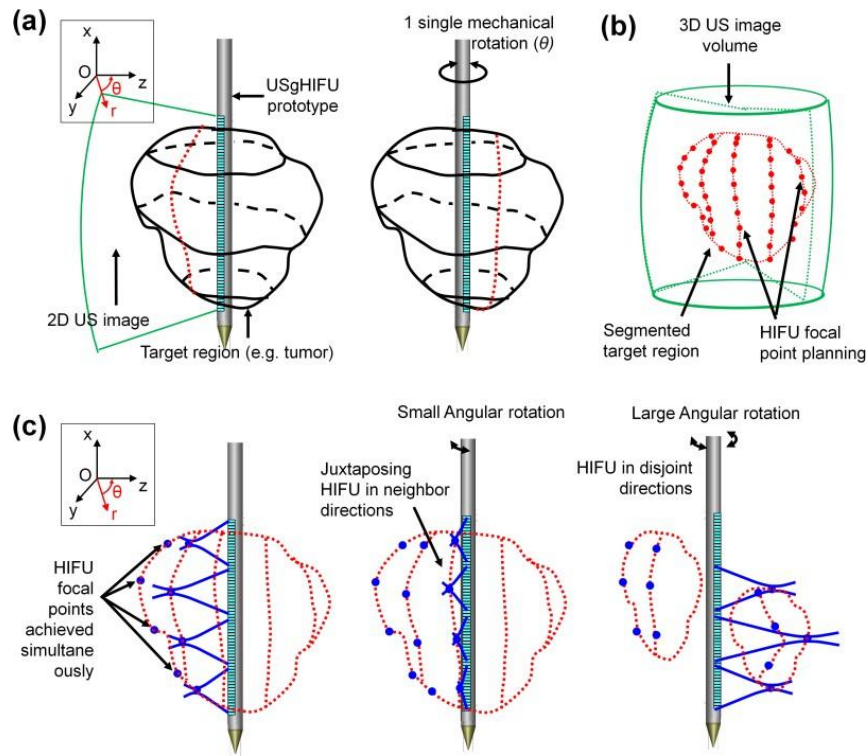

**Fig. 2**. 3D US-guided Navigation for interstitial conformal HIFU treatment planning: principle. (a) In-situ 3D US image reconstruction. A dual-mode USgHIFU interstitial prototype providing 2D sectorial B-mode US images was rotated to image the entire volume of a target region in situ (e.g. from the inside of a tumor). (b) Target region segmentation and automatic determination of the focal points for covering the 3D surface of the target region. (c) Conformal HIFU treatment. The dual-mode USgHIFU interstitial prototype providing 2D HIFU exposures with dynamic focusing capability (beam steering in the catheter elevation, z-axis) to treat a plan. Robotic assistance allowed rotating the prototype to ablate the entire volume of the target region interstitially.

### c. Dual-mode interstitial USgHIFU prototype

The dual-mode USgHIFU 9-gauge catheter prototype (Vermon, France) **(Fig. 3)** was made of a piezocomposite 64- element linear-array, 38.4 mm long and 2 mm in elevation (element pitch: 600µm; total surface area : 0.78 cm²), mounted at the tip of a rigid cylindrical rod (21 cm long, 3 mm in diameter) made of MR-compatible nitinol material. The 64 elements of the transducer were electrically independent. The central frequency $f_c$ of the array in transient wave regime (TW) was 6.0 MHz and its bandwidth @6dB was > 50%. The constant driving frequency used for both US imaging and HIFU exposures was then chosen to maximize HIFU performances in continuous wave regime (CW), as close as possible from $f_c$. At these frequencies, the ultrasound wavelength was ~250 µm, almost half the pitch of the prototype. In therapy mode (CW regime), the targeted acoustic intensities ranged 10-20 W·cm$^{-2}$. A compromise between imaging (TW regime) and therapy performances was made by not covering the transducer with any acoustic lens in order to maximize the output acoustic intensity and reduce self-heating of the device in HIFU mode.

In addition, the catheter prototype was cooled by two miniature cooling pipes (Ø = 688 µm) opening on either sides of the linear array transducer (degassed water). A thin PTFE pipe (thickness = 51 µm) was sealed over the nitinol rod at the front of the linear array to close the cooling circuit and act as an acoustic window. Water circulation was ensured using a peristaltic pump (77200-62, Cole Parmer, USA), providing a flow rate of 0.7 mL/min. The circulating water was cooled (6- 8°C) by immerging part of the external tubing in the bath of a mini-chiller (Ministat 230, Huber, Germany) containing an antifreeze liquid maintained at a temperature of -2°C. A wire thermocouple was also integrated into the catheter and placed in contact with the rear panel at the center of the transducer, allowing the temperature inside the prototype to be monitored in real-time during experiments.

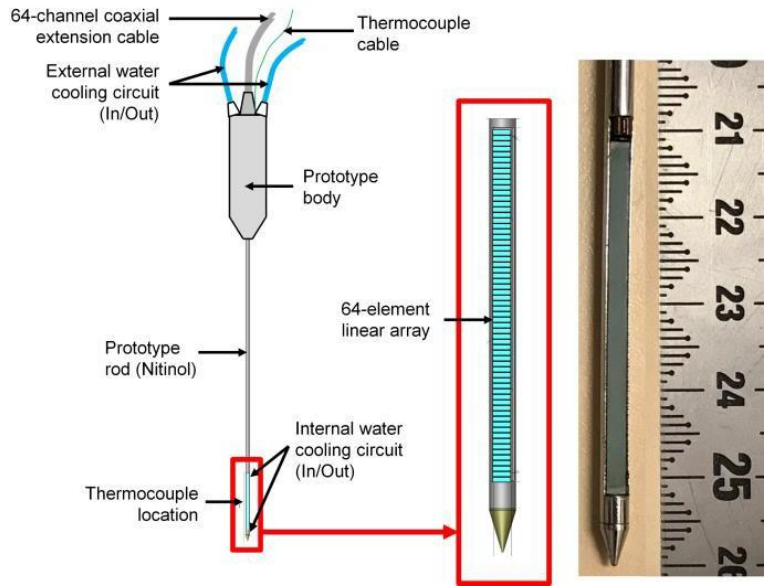

**Fig. 3**. Picture of the dual mode USgHIFU prototype. The red box zooms in on the active part containing 64 elements. A photo of the catheter is shown on the right.

### d. Dual-mode USgHIFU electrical driving chain

The dual-mode USgHIFU catheter prototype was driven by an open research ultrasound system (Vantage 256, Verasonics, United States) with 256 independent emission/reception channels. The ultrasound research platform was a dual-mode ultrasound research platform operating both as an US imaging scanner (RF pulse emitter–receiver in TW) and a HIFU driving system (high-power RF amplifier in CW). The powered CW regime was provided by associating the original US scanner hardware to an additional HIFU external power supply (QPX600DP, Aim-TTi, United Kingdom). The system was interfaced with a 1.6-m extension cable of the catheter prototype using a single 256-channel connection (UTA360 front panel, Verasonics, United States). The UTA360 connector of the extension cable integrated onboard electronic cards that performed two functions. The first function was electrical parallelization of the 256-channels dual-mode platform to minimize the level of electrical current debited on each channel and respect maximum current limitations of the generator (manufacturer data). All channels were combined four by four in parallel to form 64 electrically independent parallelized channels able to drive the 64 elements of the catheter. The second function was electrical impedance matching of the 64 parallelized channels, which was performed to maximize the electrical efficiency (electrical power transfer) in therapy mode between the HIFU generator and prototype. The onboard electronic cards provided 64 electrical pads for installing series/parallel matching circuits.

The dual-mode driving platform was not MR-compatible, thus all MR thermometry tests were carried out with the system outside the Faraday cage. This was done by replacing the 1.6- m extension cable by extension cables dedicated to MR experiments : i) a first 64-channel coaxial extension cable (3.0 m) connected outside the Faraday cage from the driving generator to ii) a grounded RF-penetration panel (UTA360 connector) fixed to the Faraday cage, and connected inside the Faraday cage (magnet room) to iii) a second 64-channel coaxial extension cable (6.0 m) directly connected to the catheter prototype, completing the driving chain within the MRI (**Fig.4**).

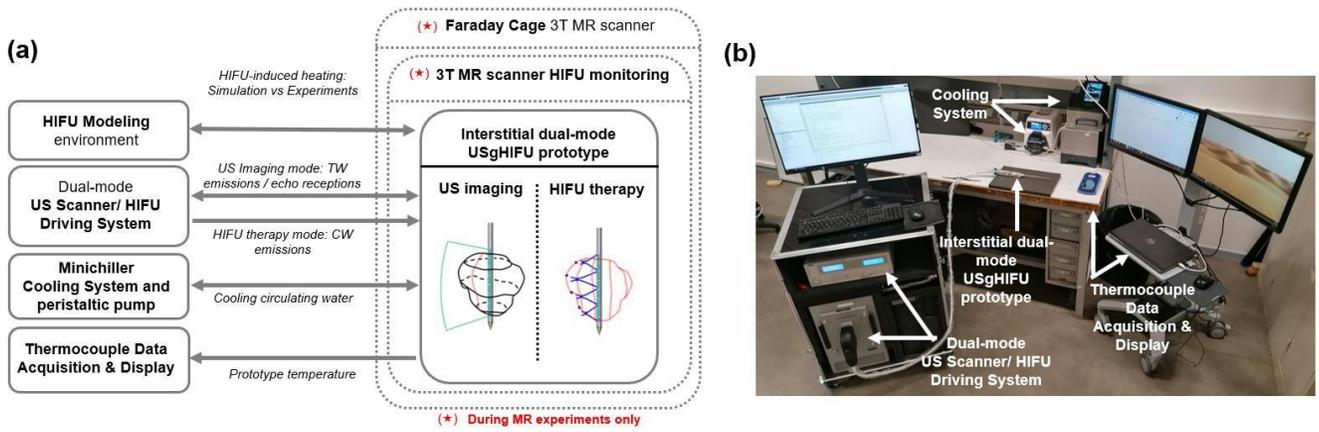

**Fig. 4**. US-image based Navigation Platform for USgHIFU therapy. (a) Schematic diagram of the communication logic between each component of the platform. (b) Photo of the open research experimental platform.

### e. Working frequency in HIFU mode: preliminary thermo-acoustic characterization

To optimize the performances of the catheter prototype in HIFU mode, a range of ultrasound working frequencies around the frequency 5.5 ± 0.5 MHz and various pulse repetition frequencies (*PRF* ranging 5-100 kHz) were preliminary explored. Especially, the ratio between the ultrasound power output and the rate of transducer self-heating had to be maximized. This ratio, referred in the manuscript as the thermo- acoustic efficiency, was defined following **(1)**:

$$\eta_{TH|AC} = \frac{<P_{ac}>_{TA}}{(\Delta T / \Delta t)} \qquad (1)$$

with $\eta_{T|AC}$ the thermo-acoustic efficiency of the transducer (W·s·°C$^{-1}$), $<P_{ac}>_{TA}$ the time average acoustic power (W), $\Delta T$ the temperature increase of the transducer (°C) and $\Delta t$ the time duration of HIFU exposure (s).

The temperature increase of the catheter was measured using the onboard thermocouple and was acquired in real-time on an 8-channel thermocouple USB data acquisition module (TC-08, Omega, USA) connected to a laptop for real-time display of the temperature curve (PicoLog 6 software, PicoTechnology, United-Kingdom). The acoustic power of the prototype was measured using the acoustic radiation force balance measurement method [40], and using a precision balance (CPA 64, Sartorius AG, Göttingen, Germany) on which an absorber (Aptflex F28, Precision Acoustic Ltd., Dorchester, UK) was suspended. The catheter prototype was inserted from the side of a tank through a cable gland and placed parallel to the absorber surface. The tank was filled with degassed water at room temperature (T = 22°C), to immerse both the catheter and absorber. The time delays applied to drive the 64 elements were set to 0 s to produce a plane wave emission (*x* axis). To preserve the prototype, these preliminary characterizations were performed at low ultrasound power. The driving voltage did not exceed 5.0 Vpp.

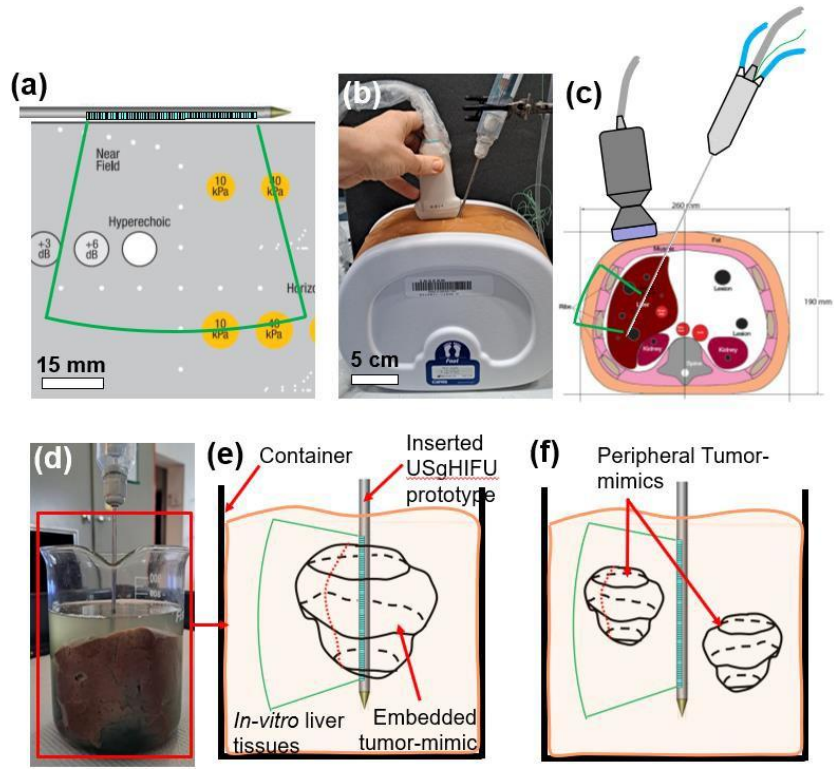

**Fig. 5**. Interstitial US imaging characterization: experimental set-ups. (a) Diagram of the imaging phantom with the catheter imaging it. (b) Insertion of the prototype into the abdominal phantom and locating the catheter using an external ultrasound probe. (c) Insertion of the protype into the abdominal phantom and visualization of the inside of the mimic organs. (d) Interstitial prototype inserted into a liver in vitro. (e) Schematic of the reconstruction of a 3-dimensional ultrasound tumor volume with the catheter inserted in a tumor mimic placed in an in vitro liver. (f) Schematic diagram of 3-dimensional ultrasound tumor volume reconstruction with the catheter inserted between two mimic tumors placed in a liver in vitro.

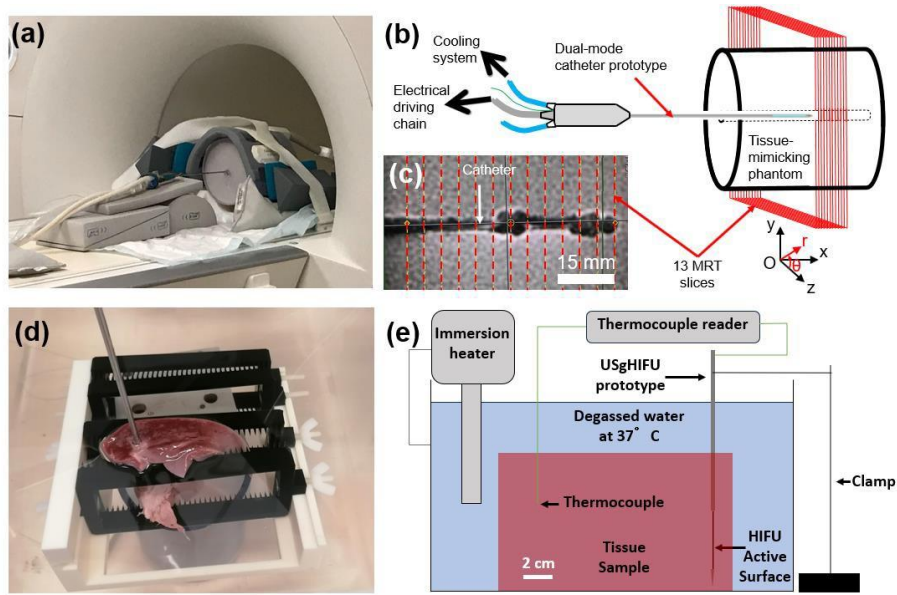

**Fig. 6**. Interstitial HIFU set-ups. (a) Experimental set-up assembled on a 3T MRI. (b) Placement of the multi-planar MRT slices axially in the azimuth plane (YZ) along the prototype in order to visualize directional heating patterns. (c) MR view in the elevation plane (XZ) of the prototype (T1- weighted 3D MP-RAGE sequence). (d) Photo of in vitro HIFU experiment. The prototype is inserted into a heifer liver. (e) Schematic of in vitro HIFU experiment.

### f. Electrical power transfer in HIFU mode: electromechanical characterization

After selecting the working frequency and *PRF* maximizing the thermo-acoustic efficiency of the transducer, impedance matching was implemented (combination of series inductor and parallel capacitor) to maximize the electrical efficiency of the driving chain defined as the ratio between the electrical power dissipated in the transducer and the total electrical power consumed by the power generator (dual-mode platform), following **(2)**:

$$\eta_{EL|EL} = \frac{<P_{el\_Cat}>_{TA}}{<P_{el\_Cat} + P_{el\_Int}>_{TA}} \quad (2)$$

with $\eta_{EL|EL}$ the electrical efficiency (%), $<P_{el\_Cat}>_{TA}$ the time average electrical power dissipated in the catheter load including the transducer and internal electronics (W), $<P_{el\_Cat} + P_{el\_Int}>_{TA}$ the total electrical power consumed by the power generator (W), $P_{el\_Int}$ being the electrical power dissipated in its equivalent internal load.

The electrical impedances $Z_{elt}$ of the 64 transducer elements were measured using a vector network analyzer (Bode 100, Omicron Lab, Austria), both in air and in water over a frequency range 2-13 MHz.

### g. Ultrasound imaging capabilities

First, the imaging capabilities of the catheter prototype were evaluated in terms of spatial resolution and contrast on a calibrated phantom (040GSE, CIRS, USA) (**Fig. 5a**). Second, the catheter was inserted into an abdominal phantom (057A, CIRS, USA) under external ultrasound guidance using an additional US imaging probe (**Fig. 5b-c**). Third, a 35 mm diameter tumor mimic was prepared in the laboratory [41] and inserted into a degassed *in vitro* heifer liver placed in a water-filled container (**Fig. 5d**). The catheter was then inserted into the tumor mimic and B-mode imaging was performed (**Fig. 5e**). Finally, two smaller diameter tumor mimic (20 and 30mm) were included into another heifer liver. The prototype was inserted between the two tumors mimic at a distance of 1 cm of each one (**Fig. 5f**). A line by line US imaging pulse echo sequence (128 emission/reception lines) was used. All 64 elements transmitted and receive at a 5.5 MHz frequency. The image obtained was a sectorial image (angular opening: 32°).

### h. Ultrasound dynamic focusing capabilities

After confirming the US imaging capabilities of the catheter prototype, its dynamic focusing capabilities were evaluated for HIFU operation in the context of achieving conformal ablations of tumor volumes exhibiting maximum radii of 30 mm. The spatial distribution of the ultrasound field was determined using the hydrophone measurement method and raster scans. Measurements were performed on a dedicated commercial platform (UMS3, Precision Acoustics, United Kingdom) including a tank (degassed water at a room temperature: 22°C), motorization enabling displacement of the hydrophone in all 3 spatial axes and a desktop computer for data acquisition and processing. A hydrophone (HGL-0200, Onda Corp, USA) was used, which exhibiting a sensitivity of 50nV/Pa and an operating frequency range 0.25–40.0 MHz. The signal acquired by the hydrophone was pre-amplified (AH-2010, Onda Corp., USA), monitored using an oscilloscope (WaveSurfer 24XS, Teledyne Lacroy, USA) synchronized with the acquisition/processing platform and the ultrasound driving generator.

For these characterizations, the transducer was driven with AC voltage pulses of 55 ultrasound cycles at 5.5MHz, a *PRF* = 1 kHz and a duty cycle of *DC* = 1%. The pressure fields spatial distributions were modulated using electronic dynamic focusing to investigate target radii, and multi-focusing vs mono-focusing performances. Raster scan acquisitions were taken with 0.3-mm spatial steps in the transducer length-elevation dimensions (*xy* plane, perpendicular to the US propagation direction) and with a 0.3-mm step in the radial dimension (*z* or *r* direction, parallel to the US propagation direction). Pressure fields were then displayed using the Paraview software (Kitware, USA).

### i. High intensity ultrasound capabilities

The ability of the catheter prototype to generate high acoustic intensities compatible with HIFU ablations was evaluated using the acoustic radiation force balance measurement method [40]. The set-up was similar to that described for preliminary characterizations, except that the driving voltage was significantly increased up to a maximum of 22 Vpp. A rather conservative maximum acceptable transducer temperature has been arbitrary set to $T_{max}$ = 45°C (based on previous prototyping) to prevent premature irreversible damage. In particular, the spatial average time average and pulse average acoustic intensities $<I_{ac}>_{SATA}$ and $<I_{ac}>_{SAPA}$ output at the transducer surface $S$ were estimated, with correspondence provided in **(3)**:

$$<I_{ac}>_{SATA} = <I_{ac}>_{SAPA} \cdot DC = \frac{<P_{ac}>_{TA}}{S} \quad (3)$$

with $DC$ the ultrasound sequence duty cycle (%) and $<P_{ac}>_{TA}$ the time average acoustic power. The electro-acoustic efficiency $\eta_{EL|AC}$ of the catheter was also evaluated to consider the global performances of the system "transducer + internal electronics", following **(4)**:

$$\eta_{EL|AC} = \frac{<P_{ac}>_{TA}}{<P_{el\_Cat}>_{TA}} \tag{4}$$

The maximum HIFU intensities which could be generated under this temperature limitation were then used for proving the concept of directional HIFU-induced: **i)** heating in tissue gel phantoms; **ii)** thermal ablations in numerically simulated tissues; **iii)** thermal ablations in *in-vitro* liver tissues. To evaluate the short- and mid-term robustness of the prototype in HIFU mode, a follow-up of the transducer performances was carried-out after each HIFU experimental session. Special attentions were paid on the generable acoustic intensity and associated transducer self-heating.

### j. Directional HIFU-induced heating in target tissues

To evaluate experimentally the capabilities of the prototype to generate directional HIFU-induced heating in target tissues, experimental sessions were first conducted in tissue phantoms and under MR Thermometry (MRT) monitoring [42,43]. This non-invasive monitoring strategy was chosen in order to access the spatial-temporal distribution of the HIFU-induced thermal heating in 3D tissues in real-time without interfering with HIFU exposures. All MR experiments were conducted using an MRI scanner (*B0* = 3.0T, Prisma, Magnetom, Siemens, Germany) hosted at the "Cermep – Imagerie du Vivant" platform. All MRT-monitored HIFU-induced heating trials were performed using a homogeneous cylindrical (Ø: 15cm, Height: 11.5cm) polyacrylamide-based gel tissue phantom (Zerdine, CIRS Inc, Norfolk, VA) with the following properties listed in **Table 1** (CIRS datasheet). These properties are similar to those reported for biological tissues in [44]. Simulations were conducted as a first approximation by neglected the effect of the variation of tissue ultrasound attenuation with temperature, thus considering a constant parameter over time.

**Table I**. Acoustic and Thermal Parameters Considered in Experiments and Numerical Modeling

| Properties | Circulating Cooling Water | Polyacrylamide Phantom | *In-vitro* Liver Tissue |
|---|---|---|---|
| Density (kg·m$^{-3}$) | 1000 | 1030 | 1050 |
| Thermal conductivity (W·m$^{-1}$·°C$^{-1}$) | 0.627 | 0.560 | 0.512 |
| Specific heat (J·kg$^{-1}$·°C$^{-1}$) | 4188 | 3800 | 3600 |
| Initial temperature (°C) | 10 | 22 | 37.2 |
| Perfusion (kg·m$^{-3}$·s$^{-1}$) | 1000 | 0 | 0 |
| Attenuation (dB·cm$^{-1}$·MHz$^{-1}$) | - | 0.5 | 0.4 |

The relative conformation of the catheter and tissue phantom was very similar to those validated and reported in previous studies of interstitial HICU/HIFU strategies [28,45,46]. The gel phantom was placed at the center of the MR scanner (**Fig. 6a**). A flexible commercial multipurpose antenna (4-channel Flex coil / Large, Siemens) was positioned around the phantom, parallel to the main magnetic field *B0*. The dual-mode USgHIFU driving research platform was positioned outside the Faraday cage and connected as described previously. The interstitial catheter prototype was inserted into the phantom, at 3.5 cm from the phantom center. The prototype was oriented parallel to the main magnetic field *B0*. Once inserted, the prototype could face > 3cm of tissue phantom radially. The distal end of the prototype was inserted over more than 40mm to immerse the entire active surface of the transducer in tissues. A MR imaging protocol previously designed and validated on the same tissue phantom model was used and adapted to the current HIFU catheter prototype and associated treatment planning configuration [28]. MRT imaging was used to measure the original tissue phantom temperature using the Proton Resonance Frequency (PRF) shift method [47–49]. A multi-planar MRT monitoring strategy consisted in the prescription of multiple 2D transverse slices, placed orthogonally to the estimated imaginary axis of insertion of the prototype in the phantom (*yz* plane) (**Fig. 6b-c**).

A series of Echo Planar Imaging (EPI) sequences associated to a parallel imaging strategy (GRAPPA) were used to increase the number of thermometry slices (2D maps) without scan time penalty [50]. The parameters implemented for the sequence are as follows: slice thickness: 2.5mm; inter slice gap: 2.3mm; bandwidth: 1250 Hz; FOV: 18.0 cm x 16.0 cm; echo time (TE): 21 ms; repetition time (TR): 832 ms. Up to 13 contiguous MR slices per second were reconstructed, covering the entire active piezocomposite part of the prototype plus peripheral tissue phantom regions. Special attentions were paid to have MRT slices coinciding with HIFU mono- and multiple-foci in order to detect the maximum temperature increases. Thermometry images were reconstructed and displayed in real-time using a modular navigation platform previously developed by our team and involving on the 3Dslicer virtual environment software (http://www.slicer.org) for the planning of US image-guided HIFU treatments [36]. Image transfer protocol and reconstruction algorithm were adapted for MR-guided HIFU treatment planning in the presented study.

### k. Directional HIFU ablations: numerical modeling

Numerical simulations of HIFU ablations were carried out prior to and in parallel with experimental investigations. HIFU field simulations were first conducted to anticipate dynamic mono-/multi-focusing performances. Second, HIFU-induced thermal heating performances were modeled in target tissues having the characteristics of the gel phantom used during MR experimental. Experimental MR thermal maps were thus used as the ground truth to tune and validate our HIFU numerical model. Especially, acoustic and thermal parameters for numerical modeling were chosen based on the literature to fit spatially and temporally experimental and simulated temperature curves for different HIFU intensity conditions. Finally, this experimentally validated numerical modeling was used to estimate theoretically the thermal ablation capabilities of the prototype.

These simulations were made using simulation software (CIVA Medical, CEA-LIST/LabTAU), an evolution of the in-house developed modeling software (ABLASIM) [51,52]. The acoustic field was calculated by solving the Rayleigh integral for a monochromatic CW for all source points of the sampled transducer surface ($\lambda/3$) [53]. All HIFU pressure fields involved in a given HIFU sequence strategy were calculated and combined sequentially over time to compute subsequent HIFU–induced thermal effects. The temperature rise was calculated by solving the Bio-Heat Transfer Equation (BHTE) with the Crank–Nicholson finite difference method [54]. The model of thermal dose based on the cumulative equivalent time at 43 °C, $t43$ in Cumulative Equivalent Minutes (CEM) was used to estimate numerically thermal damages induced by HIFU exposures in liver tissue [55]. A minimum threshold for irreversible damage was set to a commonly accepted value of 240 CEM to provide a conservative predictor of the extent of thermal ablations in liver tissues, according to hyperthermia and HIFU literature in various soft tissue types [51,56,57].

The simulated prototype was positioned to be implanted at the center of a cylindrical homogeneous target tissue. All rectangular piezoelectric elements of the transducer were modeled as homogeneous surfaces exhibiting perfect piston-like vibrations. Non-linear effects and cavitation were not considered by the software. The physical properties of the tissues and environments input for simulations are listed in **Table 1**. They were selected according to the literature and to fit simulated thermal heating to those of MR experiments [45,46]. The cooling water circulating in front of the transducer and in thermal contact with the neighbor tissues was modeled by a static water (1 mm thick) perfectly maintained at a temperature of 10°C.

### l. Directional HIFU ablation: in-vitro liver tissues

The capabilities of the catheter to generate directional HIFU ablations were evaluated in *in-vitro* liver models (pig, heifer, calf). Each liver was divided into several samples to ensure a minimal tissue thickness of 5 cm. Each tissue sample was placed and heated directly in a tank filled with degassed water and maintained at a temperature of 37°C using an immersion heater (Polystat, Thermo Fisher Scientific, USA). Liver temperature was locally monitored using a needle thermocouple (5sc-gg-k-30-72, Omega engineering, USA) inserted deep (2-3 cm) into the liver. The prototype was then inserted into the liver sample with ultrasound guidance, leaving several centimeters of tissues in front of the active HIFU transducer zone (**Fig. 6d-e**). Several treatment strategies were carried out, varying the number of prototype rotations, treatment time and type of focusing: **i)** plane wave; **ii)** mono-focal; and **iii)** quadri-focal (4 simultaneous static foci). HIFU exposures were performed using the acoustic parameters and dynamic HIFU focusing strategies included in **Table 2.** After evaluating the ability to induce directional ablations with a static catheter, more advanced targeting strategies were investigated to achieve larger directional ablations (increasing their sectorial dimensions) or to target several regions from one catheter insertion. Since the HIFU dynamic focusing was only controllable in 2D (radial & height distances), the catheter was mechanically rotated ($\theta$) around its insertion axis (*x-* or *h*-axis, to target several angular directions and cover extended volume of liver tissues in 3D.

Once the HIFU treatments were completed, liver samples were recovered for gross sample analyses. Before removing the catheter from the liver sample, its orientation was marked on the liver surface. The catheter was then carefully replaced by a stick to keep the information on its insertion trajectory *(h-*axis) and guide the upcoming liver dissection. Each liver sample was sliced with a scalpel: **i)** either along the *yz* plane to appreciate the radial and height extents of the thermal ablations**; ii)** or along the *xz* plane to appreciate the radial and angular extents (directivity) of the thermal ablations. Each tissue section was approximately 5-mm thick. All the accessible dimensions of the ablations were measured with a caliper. The ablation zone was visually assessed by observing the color change between the necrotic zone and the untreated tissues. Experimental results were also compared to numerical modeling for final validation of the modeling tool.

Table II. Dynamic HIFU Focusing Strategies

| Groups | Plane Wave | Mono-Focal Wave | Quadri-Focal Wave | Quadri-Focal Wave |
|---|---|---|---|---|
| Focal distance (mm) | - | 30 | 30 | 25 |
| Acoustic Intensity, $<I_{ac}>_{SAPA}$ (W·cm$^{-2}$) | 13.3 | 13.3 | 13.3 | 13.3 |
| Duty cycle, *DC* (%) | 70 | 70 | 70 | 70 |
| Acoustic Intensity, $<I_{ac}>_{SATA}$ (W·cm$^{-2}$) | 9.3 | 9,3 | 9,3 | 9.3 |
| Elementary HIFU Exposure duration (min) | 10 | 10 | 10; 7; 7 | 10; 7; 7 |
| Number of rotation (1) | 1 | 1 | 2 | 2 |
| Angular rotation step, $\Delta\theta$ (°) | 30 | 30 | 25 | 25 |
| Total HIFU exposure duration (min) | 20 | 20 | 24 | 24 |

### m. Data analyses

Quantitative results were presented as median value [1st quartile–3rd quartile]. Statistical significance for the comparison in the dimension of the tissue ablations between plane wave, mono- and quadri-focal groups was assessed using a Wilcoxon–Mann–Whitney test (H0: the volume of the ablations is the same between plane wave, mono- and quadri- focal groups). The significance level was set at 5%. The statistical analyses were performed using MATLAB (R2021a, The Mathworks, Inc., USA) and the BiostaTGV platform (https://biostatgv.sentiweb.fr/, UMR S 1136, INSERM, UPMC, France).

## III. RESULTS

### a. Maximizing the thermo-acoustic and electro- electrical efficiencies

Over the range of working frequencies *f* (5-6 MHz) and *PRF* (5-100 kHz) studied, the average thermo-acoustic efficiency of the catheter was of $\eta_{T|AC}$ = 0.48 ± 0.04 W·s·°C$^{-1}$ (range 0.39 – 0.55 W·s·°C$^{-1}$). Its value was maximal and equal to $\eta_{T|AC}$ = 0.55 for *f* = 5.5 MHz and *PRF* = 20 kHz (**Fig. 7g-h**). These 2 parameters were then fixed for the rest of the study, starting by the impedance matching of all 64 elements of the linear-array transducer. At 5.5-MHz frequency and before matching, the average real and imaginary parts of the element impedances in water were respectively $Re(Z_{elt})$ = 16.4 ± 2.2 Ω (range 13.7 to 20.7 Ω) and $Im(Z_{elt})$ = -65.4 ± 4.3 Ω (range -71.5 to -59.5 Ω). On the onboard internal cards, 64 LC circuits were installed (series inductor *L* = 3.1µH and parallel capacitor *C* = 1 nF) and gave average real and imaginary parts of respectively $Re(Z_{elt})$ = 38.2 ± 3.8 Ω (range 23.2 to 48.1 Ω) and $Im(Z_{elt})$ = -24.1 ± 6.0 Ω (range -41.7j to -8.0j Ω), bringing the system closer to 50Ω. All element impedance curves and catheter thermo-acoustic efficiency curves are provided in **Fig. 7**. The matching operation allowed increasing the electrical efficiency $\eta_{EL|EL}$ from 67 to 83 %. In addition to maximizing the electrical efficiency of the driving chain, the impedance matching allowed smoothing: **i)** the difference in impedance observed between 2 distinct groups of elements before matching; **ii)** the electrical voltages and currents seen by the ultrasound transducer by filtering the harmonic frequency present at the generator output. Finally, electrical parallelization of the 256- channels to drive the 64 elements changed the ratio between the internal impedance of the parallelized amplifier channels and the impedance of the prototype elements. As a result, electrical efficiency $\eta_{EL|EL}$ was increased from 83 to 95 %.

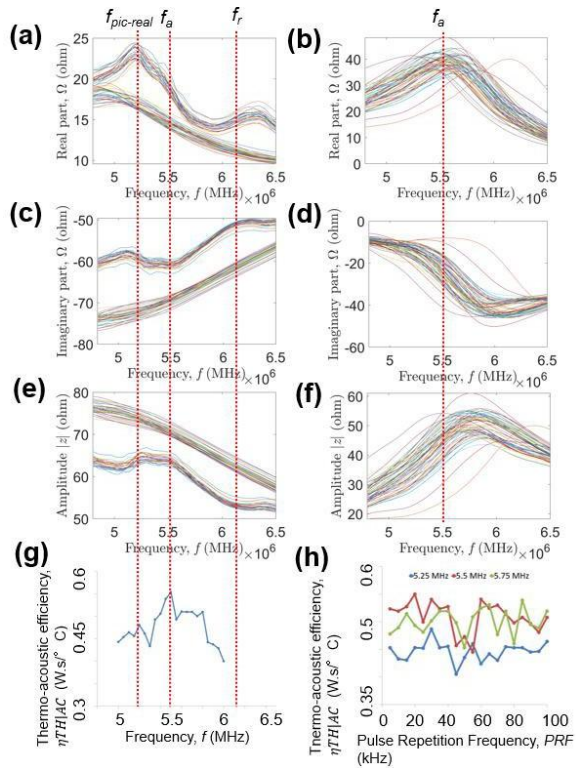

**Fig. 7**. Impedance measurements of the USgHIFU prototype. fpic-real is the frequency of the peak of the real part. fa and fr are the anti-resonance and the resonance frequencies respectively. (a) and (b): measurements of the real part before and after adaptation respectively. (c) and (d) measurements of the imaginary part before and after adaptation respectively. (e) and (f) measurements of the amplitude before and after adaptation respectively. (g) Thermo-acoustic efficiency as a function of the frequency. (h) Thermo-acoustic efficiency as a function of applied PRF for 3 operating frequencies.

### b. Real-time interstitial US imaging for HIFU guidance

For further validation of the catheter dual-modality, and although the central frequency for imaging was $f_c$ = 6MHz, all US images were acquired at a slightly lower frequency of 5.5- MHz optimized for the therapy mode (matching circuit in place). In these conditions, sectorial 2D US images could be acquired (frequency of display: 25 images/s). The field of view (FOV) extended up to 80 mm in depth ($z$ axis) and 32° in angular aperture, which corresponded to widths ($x$ axis) ranging from 38 mm (@$z$ = 0 mm) to 80 mm (@$z$ = 70 mm). Resolution was calculated using the full width at half maximum method by determining intensity peaks in regions containing inclusions and measuring their width at mid-height. The axial and lateral resolutions at the center of the image ($z$ = 30 mm & $x$ = 0 mm) were respectively of 0.55 mm ($z$ axis) and of 0.70 mm ($x$ axis) (**Fig. 8a-b)**. In the 3$^{rd}$ dimension however, the rapid ultrasound beam divergence led to a 5.8 mm transverse resolution ($y$ axis). The catheter prototype allowed clear visualization of 4 calibrated cylinder inclusions of 8 mm in diameters. Their diameters and contrasts to the surrounding tissues were retrieved with a median relative difference of 6.0 ± 4.3 % (range 3 – 14 %) and 11.0 ± 7.7 % (range 5 – 23 %) respectively (**Table 3**).

When inserted in the abdominal phantom, the catheter allowed clear visualization of the tumor mimic inclusions as well as all neighbor anatomical structures described by the manufacturer in the regions covered by the ultrasound imaging FOV. Especially, liver, tumor mimic and vein mimics were detectable from one insertion and rotation of the prototype to visualize at 360° around the catheter (**Fig. 8c-d).** Finally, similar excursions performed in *in-vitro* livers tissues confirmed the ability of the catheter to detect and precisely discriminate the boundaries of tumors-mimic inclusions: **i)** a large tumor-mimic of Ø = 35 mm (catheter inside the tumor), and **ii)** 2 smaller tumor-mimics of Ø = 20 mm and 30 mm (catheter at distance from the tumors) (**Fig.8e-h).** All quantitative results are summarized in **Table 3.**

**Table III**. Imaging Performances of the Dual-Mode USgHIFU Catheter Prototype

| Tissue Model | Element/anatomical structure | Reference values | | | Analysis on the interstitial 2D US image | | Relative difference, $\frac{|Data_{US}-Data_{Ref}|}{Data_{Ref}}$ | |
|---|---|---|---|---|---|---|---|---|
| | | Diameter, Ø (mm) | Contrast, C (dB) | Origin | Diameter, Ø (mm) | Contrast, C (dB) | Diameter, Ø (%) | Contrast, C (%) |
| Calibrated gel phantom (040GSE, CIRS, USA) | String inclusions | 8.0 | +15.0 | Manufacturer | 7.6 | +13.6 | 5 | 9 |
| | Cylindrical inclusions (tumor mimics) | 8.0 | +6.0 | Manufacturer | 6.9 | +5.2 | 14 | 13 |
| | | 8.0 | +3.0 | Manufacturer | 8.4 | +2.3 | 5 | 23 |
| | | 8.0 | -6.0 | Manufacturer | 7.5 | -5.7 | 6 | 5 |
| Abdominal gel phantom (057A, CIRS, USA) | Tumor mimics | 16.0 | - | Manufacturer | 14.0 | -14.0 | 12.5 | - |
| | Vein mimic | 0.60 | - | Manufacturer | 0.68 | -7.1 | 13.3 | - |
| *In-vitro* liver + gel tumor phantom (homemade, [41]) | Tumor mimic #1 | 25.0 | - | Gross sample | 26.5 | +24.6 | 6 | - |
| | Tumor mimic #2 | 24.0 | - | Gross sample | 23.3 | +17.5 | 3 | - |
| | Tumor mimic #3 | 12.0 | - | Gross sample | 12.5 | +18.0 | 4 | - |

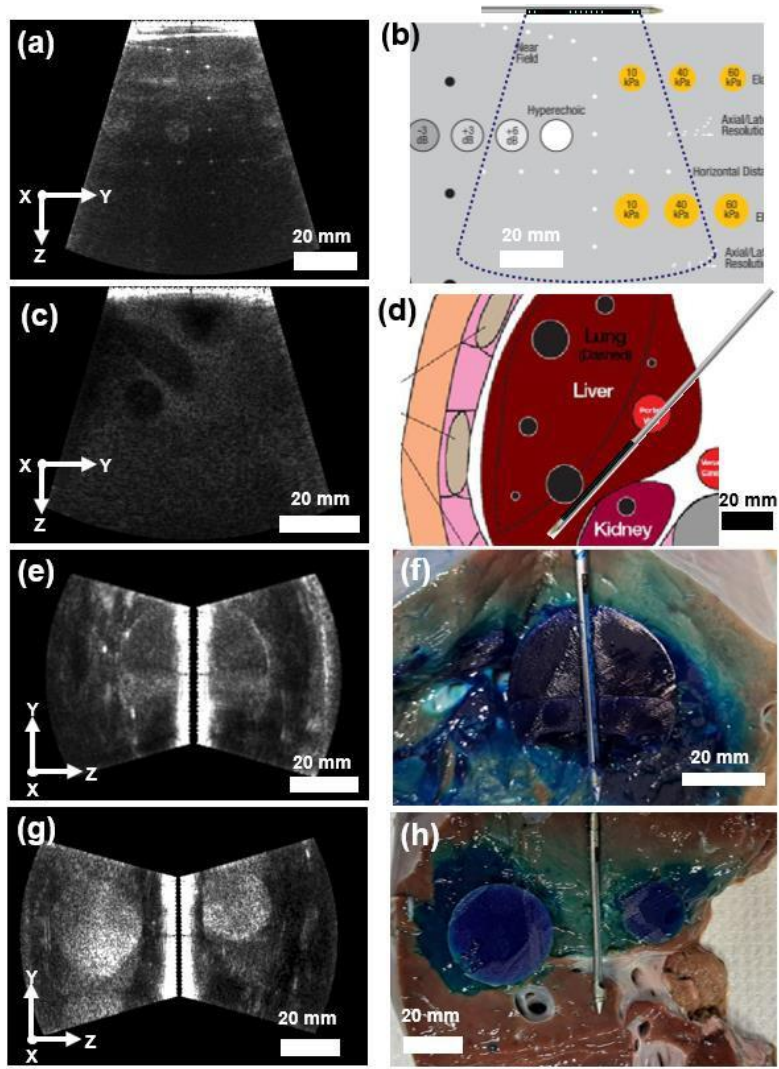

**Fig. 8**. Dual-mode ultrasound imaging. (a) Sectorial imaging of a calibrated gel phantom. (b) Region imaged by the prototype on the calibrated gel phantom. (c) Sectorial imaging of an abdominal gel phantom. (d) Region imaged by the prototype on the abdominal gel phantom. (e) Sectorial image of a tumor mimic inserted in an in vitro liver. (f) Position of the prototype inserted in the tumor mimic. (g) Sectorial image of two tumors mimic inserted in an in vitro liver. (h) Position of the prototype inserted between the two tumors mimic.

### c. Interstitial dynamic HIFU multi-focusing strategies

Several HIFU focusing strategies were investigated to optimize the directionality and spatial control of the ultrasound beam. The results are shown in **Fig. 9**. Overall, the experimental pressure fields measurements confirmed all performances anticipated during numerical modeling. Plane wave emission was associated to a low pressure and a decrease of the pressure radially (divergence of the ultrasound beam due to the very small elevation of the transducer) (**Fig. 9**).

Dynamic HIFU mono-focusing allowed to concentrate higher pressure at distance from the transducer in the controlled main lobe. However, it was also associated to the presence of significant grating lobes (- 9 dB) due to the large element pitch in relation to the wavelength. These grating lobes were not controllable since their locations varied significantly when modulating the main lobe focal distance to target various tissue radii (**Fig. 9a**).

The use of simultaneous static multi-focusing strategies, starting from bifocal focusing (two groups of 32 elements focusing on a respective single point), eliminated grating lobes. This configuration let however wide spaces (18,5 mm) with low pressures between the 2 main HIFU lobes (**Fig. 9b**). A quadri-focal strategy appeared to be a good compromise between the number of elements involved in each focusing groups (4 groups of 16 elements) and the space between the 4 main HIFU lobes (9,0 mm) (**Fig. 9d**). The quadri-focal emission produced 4 pressure peaks evenly spaced which were controllable independently, thus allowing to apply different focal distance ($z$-axis) from one focus to another in order to conform to the tissue target boundary. An example of quadri- focal emission with the outer element groups focusing at a distance of 25 mm and the two central element groups focusing at a distance of 40 mm is presented in **Fig. 9e**.

Experimental measurements were carried out at low power. Simulations at higher power levels, representative of our *in vitro* tests, were carried out. The minimum experimental and simulated pressures were 0.62 MPa and 5.3 MPa respectively. These values were obtained for a plane wave. The maximum experimental and simulated pressure was 1.60 MPa and 16.4 MPa respectively. These values were obtained for bifocal and monofocal focusing respectively.

These results showed the versatility of the prototype in focusing ultrasound energy for various targeting configurations.

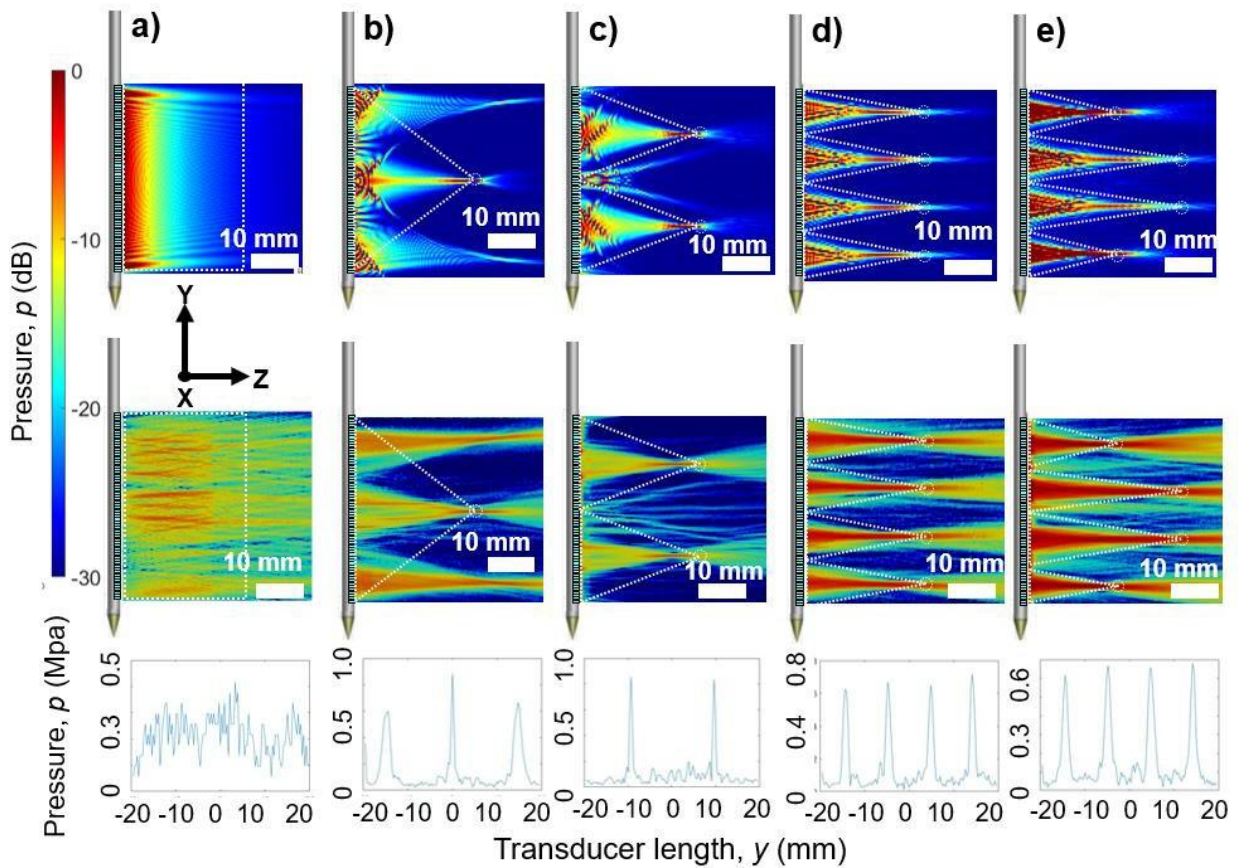

**Fig. 9**. Acoustic characterization and dynamic focusing of the prototype using pressure fields. The first line shows the simulation results, while the experimental results are shown below. Below each experimental pressure field are plotted the corresponding pressure values as a function of transducer length for a value of $z$ = 30 mm. (a) Plane wave. (b) Monofocal focusing at 30 mm. (c) Bifocal focusing at 30 mm. (d) Quadri-focal focusing at 30 mm. (e) Quadri focal, the two outermost focal points focus at 25 mm, while the two most central focus at 40 mm.

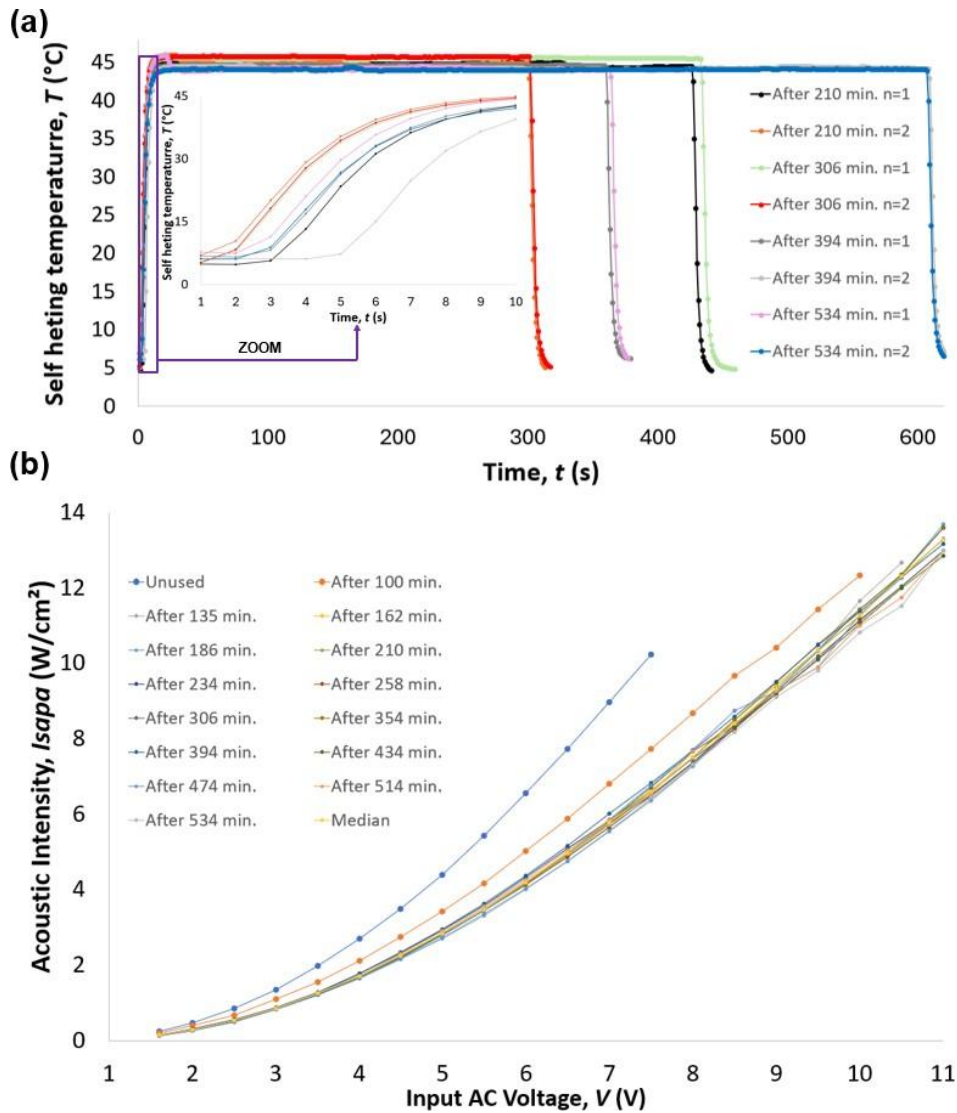

**Fig. 10**. (a) Self heating temperature of the prototype during in vitro trials. 2 layouts are made during the same in vitro test (n=1, n=2). A zoom of the temperature rise at the start of firing is shown on the right of the figure. (b) Acoustic intensity of the prototype as a function of the control voltage applied to the connector terminals. A plot was made after each use of the prototype.

### d. HIFU intensity with a 3-mm dual mode catheter

The ability of the catheter prototype to develop ultrasound energies compatible with HIFU therapies was confirmed, as well as the robustness of these performances over time (**Fig. 10b**). The maximum acoustic intensity tested robustly was $<I_{ac}>_{SAPA} = 14$ W·cm$^{-2}$ with a duty cycle of $DC = 70\%$, giving a time average intensity $<I_{ac}>_{SATA} \sim 9.8$ W·cm$^{-2}$ over the whole HIFU exposure. In total, the prototype was operated for 17 *in-vitro* trials and MRT trials corresponding to 534 cumulative minutes at full power. After a running-in period of two first trials (one of which was performed under MRI) where a decrease of 34% in the ultrasound energy performances was noticed, all performances were maintained during 400 cumulative minutes (to maintain the acoustic intensity at the set value, the control voltage was increased accordingly after the running period). In these stabilized conditions, HIFU emissions were associated to a maximum self-heating reaching a plateau at $\Delta T_{max} = 45$ °C, (**Fig. 10a**) which was in the range of temperature limit set at the begging of the study. The electro-acoustic efficiency of the catheter was $\eta_{EL|AC} = 8.4\%$.

### e. Quantitative assessment of directional HIFU-induced thermal heating in tissues

The catheter prototype was first confirmed to be MR compatible in the 3T environment. Magnetic susceptibility artifacts were observed on anatomical and MRT images, and up to 13 mm in front of the transducer on some slices could not be quantified. Real-time multi-planar thermometry monitoring of HIFU-induced thermal heating were then successfully performed over a FOV (18 cm × 16 cm × 6 cm) compatible with the monitoring of conformal HIFU treatments of large tumor volumes. HIFU-induced thermal heating could be generated by the miniaturized catheter prototype, despite the small transducer elevation and intense internal cooling (**Fig. 11**). MR experiments conducted at low- ($<I_{ac}>_{SATA}$ = 7.4 W·cm$^{-2}$) and mid- ($<I_{ac}>_{SATA}$ = 9.0 W·cm$^{-2}$) acoustic intensities revealed tissue heating extending radially over 30.5 mm and 34.5 mm respectively, after 9 min. and 10 min. respectively of HIFU exposures. In these exposure conditions, the maximum temperatures were located radially in the near field (~14 mm) and reached $\Delta T_{max}$ = 28°C and 36°C respectively. Heating was observed on all slices containing the transducer's active surface. No heating was measured behind the active surface of the transducer, confirming the directional aspect of the treatment. No significant heating was observed on sections without piezoelectric elements, indicating relatively low thermal diffusion in this direction.

These results correlated well with those obtained in numerical modeling, with a difference between confined within 4° C (or 10%) (**Fig. 11**). Further numerical modeling allowed anticipating the ability to generate HIFU ablations when using the catheter prototype at maximum acoustic intensities ($<I_{ac}>_{SATA}$ = 14 W·cm$^{-2}$). Directional HIFU ablations could be obtained in modeling which extended homogenously in the *XZ* plan, over a maximum radius (*Z* or *r* axes) of 28 mm, a height (*X* axis) of 40 mm and a thickness/angular opening (*Y* axis/$\theta$ angle) of 13 mm / 30°.

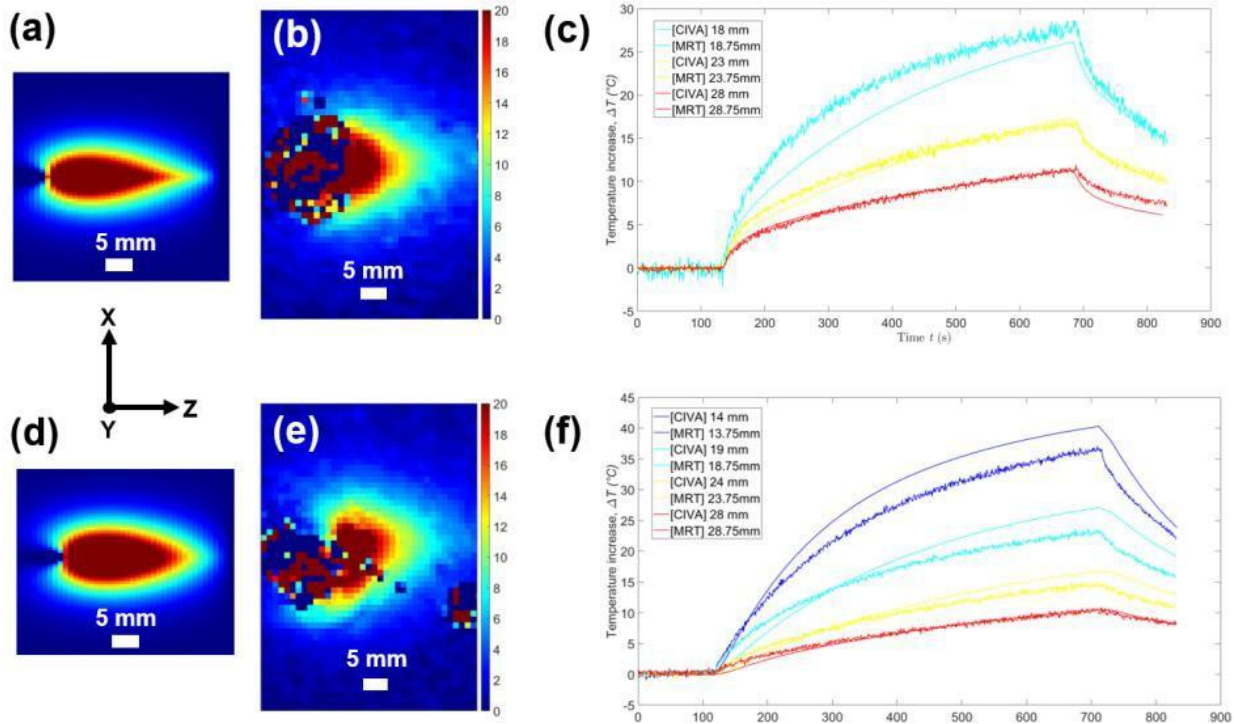

**Fig. 11**. HIFU induced directional thermal heating. (a) Numerical modeling and (b) Experimental temperature increase generated during 540s HIFU exposure (Isata = 7.4 W/cm²). (c) Comparison between simulation (CIVA) and experimentation (MRT) of temperature increase over time for 4 points distant from the prototype. (d) Numerical modeling and (e) Experimental temperature increase generated during 600s HIFU exposure (Isata = 9.0 W/cm²). (f) Comparison between simulation (CIVA) and experimentation (MRT) of temperature increase over time for 4 points distant from the prototype.

### f. Multi-directional, radial and centimetric thermal ablations in liver tissues

Last experiments allowed confirming *in-vitro* the ability to generate USgHIFU-induced thermal ablations in liver tissues. The 1$^{st}$ phase of catheter insertion within liver tissues was perfectly guided under external ultrasound imaging. The catheter was visible on external B-mode images, which allowed locating it within the liver sample, position and orientate it for tissue ablation in a given direction. The 2$^{nd}$ phase of HIFU exposures conducted using conservative acoustic intensity ($<I_{ac}>_{SATA}$ = 9.3 W/cm²), allowed robust repetitions of HIFU trials, without notable decrease in HIFU exposures intensities. Consequently, a total of 17 *in vitro* trials were successfully completed, testing 4 different strategies of ultrasound focusing: **i)** plane **ii)** mono-focal **iii)** quadri-focal waves focusing at 25 mm and **iv)** quadri-focal waves focusing at 30 mm.

A first 10-min exposure was followed by 1 or 2 mid- treatment catheter rotations for additional 7-min exposures (per additional rotation). Directional radial ablations were retrieved on gross sample analyses. HIFU ablations extended homogenously in front of the transducer in the *XZ* plan, over a median radius (*Z* or *r* axes) of 21,5 ± 3.1 mm (range 18

– 27 mm) and a height (*X* axis) of 38,0 ± 0.96 mm (range 37 – 40 mm). Rotations with a maximum angular step of $\Delta\theta$ = 30° allowed juxtaposing several directional radial ablations to form larger homogenous sectorial ablations increasing the thickness/angular opening (*Y* axis/$\theta$ angle) to 20 mm / 45°. When rotating the catheter prototype using angular steps $\Delta\theta$ > 60°, it was possible to generate independent directional radial ablations in various directions, thus to treat different regions from one insertion site. An example of catheter rotation over $\Delta\theta$ = 70° is provided in **Fig. 12b**. Both ablations were clearly visible, showing the multi-directionality of the ablative intervention. No visible tissue damages were found elsewhere around the catheter. During HIFU exposures, the catheter internal temperature did not exceed 45°C. As the p-values between each group are greater than 0.05, ablation volumes are not significantly different between each focusing approaches tested for these conservative conditions of HIFU intensities (**Fig. 13**).

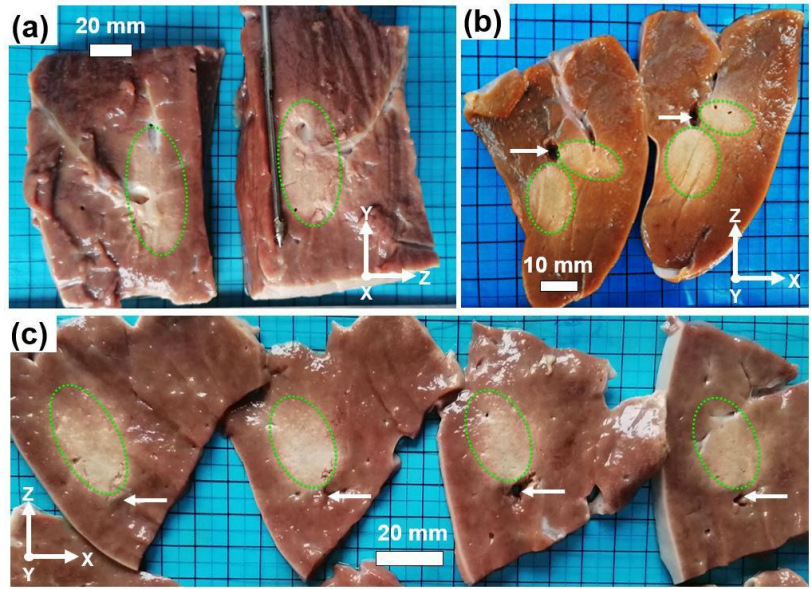

**Fig. 12**. HIFU induced (multi-) directional thermal ablations. Treated areas are circled in dotted lines. The prototype is inserted at the tip of the arrows (a) Cutting along the YZ plan. (b) Two rotations were performed at an angle of 70°. Cutting along the XZ plane. (c) Two rotations were performed at an angle of 30°. Cutting along the XZ plane.

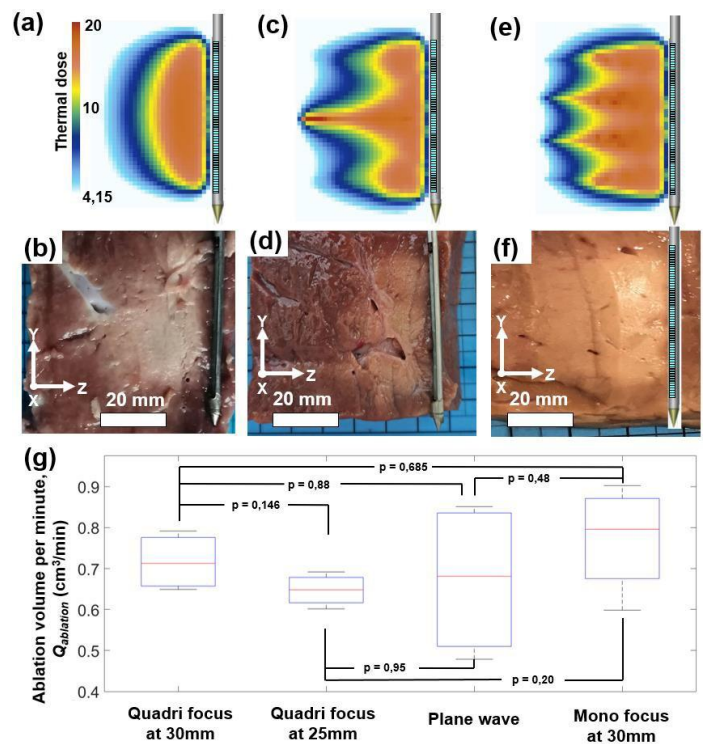

**Fig. 13**. (a-b) Simulation and experimental results for ablation with plane wave. (c-d) Simulation and experimental results for ablation with mono focusing (e-f) Simulation and experimental results for ablation with quadri-focusing. (g) Graph showing ablation volume per minute as a function of transducer focusing types.

Table IV. Comparison with the Literature

| Groups | Diameter (mm) | Active surface (mm x mm) | Number of elements | Acoustic intensity (W/cm²) | Maximal radial ablation distance (mm) | Dual mode | Strategies of use |
|---|---|---|---|---|---|---|---|
| **Academic:** | | | | | | | |
| Lyon [25][29][63][31] | 3.8 | 3 x 10 | 1 | 14 to 20 | 11 | No | Interstitial |
|  | 3.5 | 2.5 x 7.5 | 1 | 11 to 17 | 10 | Yes* | Interstitial |
|  | 4.0 | 3 x 10 | 1 | 30 to 55 | 19.8 | No | Interstitial |
|  | 3.8 | 3 x 20 | 5 | 18 | 15 | Yes* | Interstitial |
| Cincinnati [32][33] | 3.3 | 2.3 x 49 | 32 | > 80 | 35 | Yes | Interstitial |
|  | 3.1 | 2.3 x 20 | 32 | > 110 | 21.5 | Yes | Interstitial |
| San Francisco [35] | 9.0 | - | 32-128 | 10.45 | 10 | No | Interstitial |
| Toronto [18][45] | 4.0 | 4 x 15 | 1 | - 20 | 10 | No | Endo-urethral |
|  | - | - | 4-5 |  | 31 | No | Endo-urethral |
| Prototype presented in the study | 3.0 | 2 x 38 | 64 | 14 | 28 | Yes | Interstitial |
| **Commercial:** | | | | | | | |
| Tulsa Pro [62] | 7.5 | ~5 x 50 | 10 | - | 30 | No | Endo-urethral |

*With mechanical rotation

## IV. DISCUSSION

In this study, we have shown that it was possible to perform *in vitro* (multi-)directional ablations using a miniature ultrasound catheter guided by integrated imaging.

### a. Interest of this catheter design for interstitial USgHIFU

Compromises were made in the design of this catheter to optimize performance for image-guided interstitial HIFU therapy. The linear-array of this V2 catheter was long enough (38 mm) to cover heights similar to those of large HCCs in a single insertion, without the need for translation - an improvement on the V1 catheter previously developed [36]. The catheter pitch was greater than the wavelength, but the resulting imaging was of sufficient quality to detect targets such as tumor mimics. Real-time 2D imaging provided additional information compared with other prototypes developed previously [30,31]. The prototype was compatible with the 3D ultrasound navigation platform while now enabling experimentally HIFU ablations. The thermo-electrical efficiency was maximized for a frequency (5.5 MHz) between the anti-resonance and resonance frequencies. In these conditions, high surface intensities comparable to those generated with larger transducers were generated robustly, despite the increased electrical and mechanical constrains associated to miniaturization. The different focusing strategies made it possible to overcome grating lobes and were made possible by the specific configuration of this catheter. The ability of ultrasound to reach deep into tissue and integrated cooling allowed for directional radial treatments, creating or expanding independent ablation volumes. All this was achieved with a single needle insertion, without any trace of unwanted damage to surrounding tissue. The internal cooling system first ensured the robustness of the prototype and maintained its performance over time. But cooling also allowed protecting non-targeted tissues from uncontrolled omnidirectional thermal diffusion effects, having a significant impact given current HIFU exposure times. This dual mode USgHIFU catheter remains, to our knowledge, the thinnest of this kind in current literature (<9G), with dimensions similar to some standard devices in the field of interventional radiology and NOTES. A comparison of the prototype with some catheters from the literature is shown in **Table 4.**

### b. Interest of this dual-mode catheter in interventional radiology

Although compatible with interstitial applications, the diameter of this prototype (3 mm) represented a maximum acceptable dimension in interventional radiology. This aspect can however be balanced by the capacity of the prototype to provide an in-situ US B-mode image from the heart of the target (e.g. HCC tumor) in real-time, a modality not currently available with other interstitial tools involved in interventional radiology. This modality can be used to i) guide the practitioner and target an area for treatment; ii) characterize target tissues; iii) plan conformal treatments [36] and iv) monitor ablations (elastography, ultrasound thermometry [58,59]), the catheter being compatible with ultrafast imaging. The development of such USgHIFU catheter in conjunction with 3D ultrasound navigation, fusion imaging (US/CT) and medical robotics should contribute to move the procedure towards personalized therapies, adapting to the shape, type and environment of the target to be treated. This interstitial US navigation-guided conformal HIFU strategy being MRI-independent could be adapted to medical services without MRI. However, the V2 catheter became MR-compatible (contrary to the V1) which allow enlarging its potential use to other medical indications where MRI is part of the procedure (guidance/monitoring). In our study, this MR compatibility allowed using MRI as a metrological tool for validating HIFU numerical modeling tools and characterize in real-time 3D the HIFU performances of the prototype, which is mandatory to further develop more advanced conformal therapies. All the advantages presented by this new USgHIFU interstitial catheter prototype (conformational potential, in-situ guidance via dual-mode imaging, possible coupling with robotics and 3D navigation, compatibility with MRI monitoring) mean that it could also be of interest in other medical indications where current interstitial techniques are promising but still present limitations for access to conformal 3D treatment of the target.

### c. Remaining technical challenges

The main challenges are the miniaturization of the device and the electronic integration required. The electroacoustic characteristics of piezoelectric technology also have some limitations. Loss of performance of mechanical and thermal origin and fragility (transducer depolarization, deformation) are challenges to be met. The ultrasound working frequency is also limited in therapy, even for transducers considered broadband in imaging. Dual mode HIFU control platforms are complex and expensive and marketing may require significant financial resources. Finally, the chain of command for miniaturized devices is also complex. It is necessary to maximize energy transfer and electro-acoustic transformation on highly mechanically and electrically constrained systems. The low electro-acoustic efficiency is explained by the miniature electronic and mechanical integrations coupled to a material whose electro-acoustic efficiency alone (without control chain or internal electronics) was 40%. The tests presented in this study showed the robustness of the prototype on average power, making it possible to validate the acoustic performance. It is now a matter of increasing the intensity delivered by the prototype and verifying that it remains robust at higher intensities. Ablations will then be generated much more quickly.

### d. Anticipated potential of interstitial dual-mode USgHIFU therapies

In this study, particular attention was given to the demonstration of the feasibility but also of the repeatability of the tests, in particular by reporting the monitoring of performance over several months of HIFU experiments. In our opinion, these elements are crucial to hope to see the emergence of interstitial approaches in the clinic. For this first proof of concept, a complete study could be carried out because we chose to validate conservative parameters. Nevertheless, the numerical results suggest even more interesting performances (notably in processing time), but not only (maximum radii). The treatment time may be interesting to improve but the current performances are not blocking, given the procedure times associated with the interventions between preparation of the patient, guidance imaging, insertion of one or of several needles (e.g. RFA), treatment planning, post-treatment control [60,61]. Alternative MEMS technologies (Capacitive and Piezoelectric Micromachined Ultrasound Transducers: CMUT and PMUT) are under development and would make it possible to overcome certain current limitations (bandwidth, miniaturization, integration density [62]), but challenges still need to be solved in order to achieve power performance levels similar to conventional bulk piezoelectric technologies. Dual- mode platforms developed in association with other systems (robotics, optical or magnetic tracking) would make the system more versatile. Finally, the *in-situ* ultrasound imaging modality would allow the integration of other imaging methods (thermometry, elastography, perfusion Doppler, contrast imaging) for the characterization of tumors, structures to be protected or even the monitoring of ablations.

## V. CONCLUSION

In this work, we have demonstrated the feasibility of centimeter-scale and directional *in vitro* thermal ablations using a 3 mm diameter ultrasound bimodal interstitial catheter. The study of different dynamic focusing strategies overcomes the physical problems associated with the miniaturization of the device. The prototype was able to B-mode image in real time and detect targets such as tumors mimic. The depth and intensity of the thermal ablations performed are encouraging and open a wide range of tumors treatment prospects. The results of this study show that interstitial ultrasound catheters may represent a promising new approach in this field.

## VI. ACKNOWLEDGEMENTS

(ANR-11-IDEX- 0007) operated by the French National Research Agency (ANR).

The authors would like to thank the team members involved at Vermon: Mathieu LEGROS, Stéphane LAFON, and Marie DENEAU for their involvement in the development of the interstitial prototype The authors would also like to thank the CERMEP team, especially Dr. Franck LAMBERTON and Dr. Danielle IBARROLA for their helpful assistance in the setup of MR-sequences. Finally, the authors would like to thank the members of the LabTAU team (INSERM U1032) who have been involved in this project, especially Dr. Françoise CHAVRIER and Victor DELATTRE for their helps in the use of CIVA HIFU modeling software.

## VIII. BIOGRAPHIES

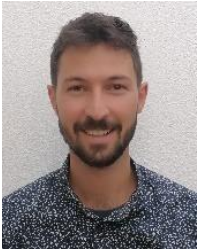

**Thomas Biscaldi** was born in Hyères, France, in 1998. He received an M.Eng. degree in biomedical engineering in Lyon in 2021. He then joined in 2021 the French National Institute of Health and Medical Research, INSERM U1032, Lyon, to start a thesis on the development of an ultrasound interstitial prototype for the treatment of hepatocellular carcinoma. He has also passed the French agrégation of electrical engineering, a highly competitive exam for teaching in higher education.

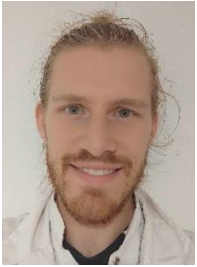

**Romain L'Huillier** was born in Nancy, France in 1992. He received an MD. Degree in radiology and an M. Eng degree in biomedical engineering from the University Claude Bernard Lyon I (UCBL, University of Lyon, France) in 2021 and 2022 respectively. Since 2022, he has been completing a PhD in science at the French National Institute of Health and Medical Research, INSERM U1032, Lyon on the development of ultrasound interstitial prototype for the treatment of hepatocellular carcinoma. He is currently a diagnostic and interventional radiologist at University Hospital of Lyon, France and a member of the French Comprehensive Liver Center, Hospices Civils de Lyon, University of Lyon, Lyon, France.

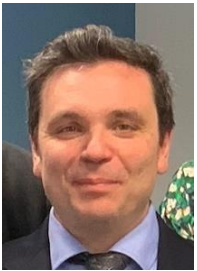

**Laurent Milot** After finishing medical school at Paris V, Pr Milot got his MD and MSc at the university of Lyon as a specialist in radiology. After a fellowship at the University of Toronto, he was subsequently hired as assistant professor in the medical imaging department and staff radiologist at Sunnybrook Health Science Center. He became associate professor, associate vice-chair of research and associate member at the Institute of Medical Sciences. After 11 years in Toronto, he came back to Lyon as deputy chief of the department of medical imaging at the Hôpital Edouard Herriot. His subspecialty is in diagnostic and interventional body radiology. After getting his PhD and HDR, he has been appointed as Full Professor of Diagnostic and Interventional Radiology in September 2021 at the University Claude Bernard Lyon 1 and Hospices Civils de Lyon and is a clinician scientist at the LabTAU, INSERM. Since then, he has been building an innovative clinical and research program linking innovations in diagnostic (radiomics, AI) and interventional (robotics, multimodal treatment, HIFU) radiology.

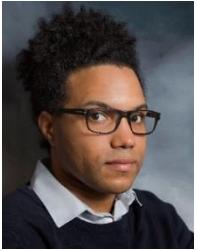

**W. Apoutou N'Djin** (Member, IEEE) received the M. Eng. degree in Instrumentation from the National Graduate School of Engineering & Research Center of Caen (Ensicaen, France) in 2004, the M.S. degree in image & signal processing from the Centrale Graduate School of Lyon (ECL, France), in 2005, and the Ph.D. degree in biomedical engineering from the University Claude Bernard Lyon I (UCBL, University of Lyon, France), in 2008.
From 2005 to 2008, he worked as a PhD Student at the Laboratory of Therapeutic Applications of Ultrasound (LabTAU, Lyon, France) of the French National Institute of Health and Medical Research (INSERM) and UCBL, on the development of intraoperative Ultrasound-guided High Intensity Focused Ultrasound (USgHIFU) therapies for focal thermal ablations of liver metastases. From 2009 to 2011, he joined the Imaging Research Laboratory of the Sunnybrook Research Institute (SRI, Toronto, Canada) and University of Toronto (U of T, Canada), as a Post-Doctoral Research Fellow to study endocavitary/interstitial MR-guided High Intensity Collimated Ultrasound (MRgHICU) therapies for conformal thermal ablations of localized tumors in prostate and brain. In 2012, he returned to LabTAU (INSERM, UCBL, Lyon, France), and has worked there since 2013 as a Research Associate and Principal Investigator. His research interests include focal/conformal image-guided HIFU/HICU therapeutic strategies, novel MEMs-based ultrasound technologies (CMUTs), and the mechanisms & applications of focused ultrasound (FUS) neuro-stimulation/-modulation. He is involved in translational research, from numerical modeling to experimental development.
Dr. N'Djin has been a member of the International Society of Therapeutic Ultrasound (ISTU) since 2011, and the International Transcranial Ultrasonic Stimulation Safety and Standards (ITRUSST) consortium since 2021. Since 2022, he has served as a Member of INSERM's national Specialized Scientific Commission "Technologies for Health". In 2019, he received the international Frederic LIZZI Early Career Award of the ISTU (ISTU/EUFUS symposium, Barcelona, Spain).